\documentclass[10pt]{article}

\usepackage[T1]{fontenc}
\usepackage[utf8]{inputenc}
\usepackage[english]{babel}
\usepackage{amsmath,latexsym,amsfonts,amsthm,mathtools,dsfont,bigints}
\usepackage{graphicx}
\usepackage[top=1.5in, bottom=1.5in, left=1.25in, right=1.25in]{geometry}
\usepackage{booktabs,url}

\usepackage{rotating}
\usepackage{tikz}
\usetikzlibrary{positioning,arrows.meta}

\usepackage{multirow}
\usepackage{color, colortbl}
\definecolor{gray}{gray}{0.9}

\newcommand{\Prob}[1]{\mathbb{P}(#1)}

\newcommand{\indep}{\raisebox{0.05em}{\rotatebox[origin=c]{90}{$\models$}}}

\newcommand{\digamma}[1]{\Psi\left(#1\right)}

\DeclareMathAlphabet{\mathcall}{OMS}{zplm}{m}{n}

\title{A probabilistic tree model to analyze fuzzy rating data}
\author{Antonio Calcagn\`{i}$^{(1)}$\footnote{Member of the INdAM research group GNCS (National Institute of Advanced Mathematics; National Group of Scientific Computing).} ~and~ Luigi Lombardi$^{(2)}$ \\\\
		\footnotesize{$^{(1)}$\sl University of Padova}, 
		\footnotesize{$^{(2)}$\sl University of Trento}\\
		\footnotesize{E-mail: antonio.calcagni@unipd.it}
	}

\date{}

\begin{document}

\maketitle

\begin{abstract}
\noindent In this contribution we provide initial findings to the problem of modeling fuzzy rating responses in a psychometric modeling context. In particular, we study a probabilistic tree model with the aim of representing the stage-wise mechanisms of direct fuzzy rating scales. A Multinomial model coupled with a mixture of Binomial distributions is adopted to model the parameters of LR-type fuzzy responses whereas a binary decision tree is used for the stage-wise rating mechanism. Parameter estimation is performed via marginal maximum likelihood approach whereas the characteristics of the proposed model are evaluated by means of an application to a real dataset.\\

\end{abstract}

\vspace{1cm}

\section{Introduction}

Rating data are ubiquitous across many disciplines that deal with the measurement of human attitudes, opinions, and sociodemographic constructs. In these cases, as the measurement process involves cognitive actors as the primary source of information, the collected data are often affected by fuzziness or imprecision. Fuzziness in rating data has multiple origins, which go from the semantic aspects of the questions/items being rated to the decision uncertainty that affects the rater response process \cite{Calcagn2021}. By and large, the differences along this continuum might reflect the differences between the ontic and epistemic viewpoint on fuzzy statistics \cite{Couso_2014}. To give an example of what is intended with fuzziness as decision uncertainty, consider the case where a rater is presented with a question/item ``I am satisfied with my current work'' and a five-point scale ranging from ``strongly disagree'' to ``strongly agree''. In order to provide a response - which corresponds to mark one of the five labels/levels of the scale - a rater behaves according to a sequential process, the first step of which consists in the opinion formation stage in which cognitive and affective information about the item being rated - i.e., job satisfaction - are retrieved and integrated until the decision stage is triggered (second step). This includes the selection stage, where the set of response choices is pruned to obtain the final rating response, for example ``strongly agree''. Decision uncertainty arises from the conflicting demands of the opinion formation stage (first step), which requires the integration of often conflicting cognitive and affective information (for instance, a work problem with the boss might increase the probability of answering the item negatively) \cite{leary1990impression}. Stated in this way, fuzziness does not reflect an ontic property of the item being rated, rather it originates from the cognitive demands underlying the response process, namely the epistemic state of the rater. 

\begin{figure}[!h]
	\centering
	\includegraphics[scale=0.6]{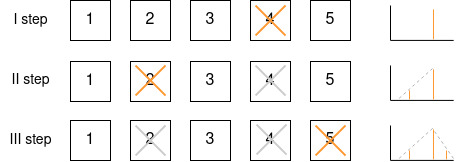}
	\caption{Example of direct fuzzy rating scale with five-point levels along with the resulting triangular fuzzy response.}
	\label{fig1}
\end{figure}

Over the recent years, a number of fuzzy rating scales have been proposed to quantify fuzziness from rating data, including both direct/indirect fuzzy rating scales and fuzzy conversion scales (for an extensive review, see \cite{calcagn2021psychometric}. In addition, see \cite{rosa2013fuzzy,de2014fuzzy,lubiano2021fuzzy} for further developments on this topic). In its most typical implementation, a (direct) fuzzy rating scale allows the rater to provide his/her response by adopting a stage-wise procedure \cite{hesketh1988application,lubiano2016descriptive}. To exemplify, consider the following five-point scale: (1) ``strongly disagree'', (2) ``disagree'', (3) ``neither agree nor disagree'', (4) ``agree'', (5) ``strongly agree''. First, the rater marks his/her choice on the scale (e.g., ``agree'') and then he/she extends the previous choice by marking another point both on the left (e.g., ``disagree'') and right (e.g., ``strongly agree'') sides. Finally, the marks are integrated to form a triangular fuzzy number where the core of the set is linked to the first mark whereas the support of the set is linked to the left and right extensions. Figure \ref{fig1} shows a graphical representation of such a procedure.

Fuzzy rating data can be analyzed either by means of standard statistical approaches or by adopting fuzzy statistical methods devoted to this purpose. In the first case, fuzzy numbers need to be turned into crisp numbers in advance through a defuzzification procedure whereas in the second case fuzzy numbers are used as is. Several fuzzy statistical methods are available nowadays (for a recent review, see \cite{couso2019fuzzy}). However, as for many statistical models, they are quite general and, in the case of fuzzy rating data, these models do not offer a thorough formal account of the mechanism underlying the fuzzy rating process. 

In this contribution, we introduce a novel statistical model to analyse LR-type triangular fuzzy data $\text{Trg}(c,l,r)$. The aim is to provide a tailor-made statistical model which mimics the stage-wise response process of direct fuzzy rating scales as those developed by \cite{hesketh1988application} and \cite{lubiano2016descriptive}. In particular, such a model would be of great interest for those who are interested in studying the relationships among fuzzy rating responses and other variables (e.g., covariates) from the perspective of the mechanisms at the origin of fuzzy responses (i.e., the three-stage response mechanism). The remainder of this short paper is organized as follows. Section 2 describes the model along with the estimation procedure. Section 3 describes the results of a real case study used to assess the features of the proposed model. Finally, Section 4 concludes this contribution by providing final remarks and suggestions for future extensions.

\section{Model}

Let $\mathbf{\tilde y} = \left( (c_1,l_1,r_1),\ldots,(c_i,l_i,r_i),\ldots,(c_I,l_I,r_I) \right)$ be a $I \times 1$ sample of triangular fuzzy numbers represented using the LR parameterization. In this context, $c_i \in \{1,\ldots,M\}$ is the core of the fuzzy number and represents the first step of the stage-wise rating process, $l_i \in \{0,\ldots,M-1\}$ is the left spread of the fuzzy number and codifies the second step of the rating process, whereas $r_i \in \{0,\ldots,M-1\}$ is the right spread of the fuzzy number and codifies the last step of the rating process ($M$ is the number of levels of the rating scale). The magnitude of $l_i$ and $r_i$ quantifies the fuzziness of the rating process. It is straightforward to notice that the data encapsulate two types of uncertainty, one related to the sampling mechanism (i.e., randomness) and one related to the response process (i.e., the decision uncertainty expressed in terms of fuzziness). We assume that fuzziness results from the interplay among different components such as the characteristics of the item/question being assessed (e.g., the easiness, with higher values being associated to less difficult items in terms of response process), the characteristics of the rater (e.g., his/her ability to respond the item), and further contextual factors like social desirability, faking or cheating. For the sake of simplicity, as in the traditional Rasch modeling framework \cite{van2016handbook}, we shall consider the first two components only, namely the item $\alpha \in \mathbb R$ and the rater's ability $\eta_i \in \mathbb R$. Under the stage-wise mechanism depicted in Figure \ref{fig1}, the probability of a fuzzy response can be factorized as follows:
\begin{align}
	\Prob{Y_i = (c,l,r)|\eta_i; \boldsymbol{\theta}} =&~ \Prob{C_i = c|\eta_i; \boldsymbol{\theta}}\cdot \label{eq1a}\\
	& \cdot \Big[ \xi_i\Prob{L_i = l|C_i,\eta_i; \boldsymbol{\theta}}\Prob{R_i = r|C_i,\eta_i; \boldsymbol{\theta}} + \label{eq1b}\\
	& + (1-\xi_i)\Prob{L_i = 0|C_i,\eta_i; \boldsymbol{\theta}}\Prob{R_i = 0|C_i,\eta_i; \boldsymbol{\theta}} \Big] \nonumber
\end{align}
where \eqref{eq1a} indicates the probability model for the first step of the rating process, \eqref{eq1b} represents the second and third steps of the rating process, $\boldsymbol{\theta}$ is a real vector of parameters which governs the behavior of the model (to be specified later), whereas $\xi \in [0,1]$ controls the mixture component of the model. Note that (i) conditionally on $C_i$, $L_i$ and $R_i$ are independent (i.e., $L_i \indep R_i$, for all $i \in \{1,\ldots,I\}$), (ii) the mixture component \eqref{eq1b} allows for disentangling those situations involving a certain level of decision uncertainty (i.e., $\xi_i > 0$) from those situations with no decision uncertainty (i.e., $\xi_i = 0$). In what follows, we will describe all the terms involved by Eqs. \eqref{eq1a}-\eqref{eq1b} in more details.

\subsection{About the probabilistic term \eqref{eq1a}} 

\begin{figure}[!h]
	\resizebox{5cm}{!}{\begin{tikzpicture}[auto,vertex1/.style={draw,circle},vertex2/.style={draw,rectangle}]
			\node[vertex1,minimum size=1cm] (eta1) {$\text{node}_1$};
			\node[vertex1,minimum size=1cm,below right= 1.5cm of eta1] (eta2) {$\text{node}_2$};
			\node[vertex1,minimum size=1cm,below right= 1.5cm of eta2] (eta3) {$\text{node}_3$};
			\node[vertex2,below left= 1.5cm of eta1] (y1) {$Y=1$};
			\node[vertex2,below left= 1.5cm of eta2] (y2) {$Y=2$};
			\node[vertex2,below right= 1.5cm of eta3] (y3) {$Y=4$};
			\node[vertex2,below left= 1.5cm of eta3] (y4) {$Y=3$};	
			\draw[->] (eta1) -- node[right=0.5 of eta1] {} (eta2);
			\draw[->] (eta2) -- node[right=0.5 of eta2] {} (eta3);
			\draw[->] (eta1) -- node[left=0.5 of eta1] {} (y1);
			\draw[->] (eta2) -- node[right=0.5 of eta2] {} (y2);
			\draw[->] (eta3) -- node[left=0.5 of eta2] {} (y3);
			\draw[->] (eta3) -- node[left=0.5 of eta3] {} (y4);			
\end{tikzpicture}}
	\hspace{2cm}
	\resizebox{5cm}{!}{\begin{tikzpicture}[auto,vertex1/.style={draw,circle},vertex2/.style={draw,rectangle}]
			\node[vertex1,minimum size=1cm] (eta1) {$\text{node}_1$};
			\node[vertex1,minimum size=1cm,below right=1cm of eta1] (eta3) {$\text{node}_{2}$};
			\node[vertex2,below left=1cm of eta1] (Y3) {$Y=3$};
			\node[vertex1,minimum size=1cm,below right=2cm of eta3] (eta3a) {$\text{node}_{4}$};
			\node[vertex1,minimum size=1cm,below left=2cm of eta3] (eta3b) {$\text{node}_{3}$};
			\node[vertex2,below left=0.5cm of eta3b] (Y1) {$Y=1$}; \node[vertex2,below right=0.5cm of eta3b] (Y2) {$Y=2$};
			\node[vertex2,below left=0.5cm of eta3a] (Y5) {$Y=4$}; \node[vertex2,below right=0.5cm of eta3a] (Y6) {$Y=5$};
			
			\draw[->] (eta1) -- node[right=0.5 of eta1] {} (eta3);
			\draw[->] (eta1) -- node[right=0.5 of eta1] {} (Y3); 
			\draw[->] (eta3) -- node[right=0.5 of eta1] {} (eta3a);
			\draw[->] (eta3) -- node[right=0.5 of eta1] {} (eta3b);
			\draw[->] (eta3b) -- node[right=0.5 of eta1] {} (Y1); \draw[->] (eta3b) -- node[right=0.5 of eta1] {} (Y2);
			\draw[->] (eta3a) -- node[right=0.5 of eta1] {} (Y5); \draw[->] (eta3a) -- node[right=0.5 of eta1] {} (Y6);
			\end{tikzpicture}}
	\caption{Examples of IRTree models for modeling response processes in rating scales.}
	\label{fig2}	
\end{figure}
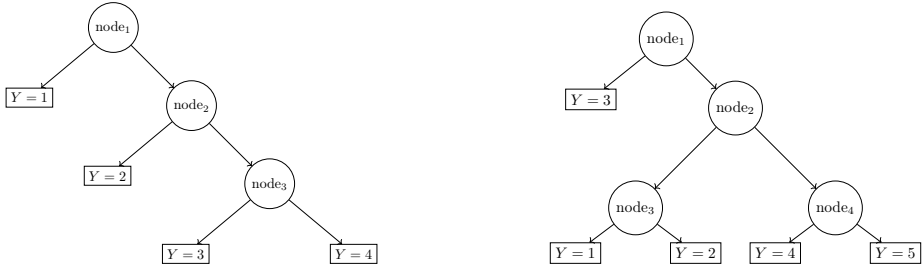

To instantiate the first term of the joint probabilistic model, we use the Rasch-tree model which is part of the family of IRTrees \cite{Boeck_2012,van2016handbook}. Among other advantages, they offer a simple and effective statistical representation of rating responses in terms of conditional binary trees \cite{Boeck_2012,Jeon_2015}. Figure \ref{fig2} shows two examples of IRTree for modeling rating responses. 

\noindent More formally, we set:
\begin{align}
	\Prob{C_i = c|\eta_i; \boldsymbol{\theta}} = \mathcall Multinom(\mathbf c^\dagger; 1,\boldsymbol{\pi}_i^y) \tag{1.1}
\end{align} 
where $\mathbf c^\dagger_i \in \{0,1\}^M$ is the event $C_i=c$ represented as a Boolean vector via the indicator function $\mathcall I (C_i^\dagger = c )$. {Note that, in light of the mapping between Multinomial and Categorical random variables \cite{murphy2012machine}, the outcomes of $C_i$ can be rewritten using a dummy vector with $M$ elements, all of which are zero except for the entry $C_i=c$. For example, the event $C_i=3$ can be rewritten as $c^\dagger=(0,0,1,0,0)$. The $M\times 1$ vector of probabilities $\boldsymbol{\pi}^y_i$ is defined according to a user-defined IRTree model as follows:
	\begin{align}
		& \pi^y_{im} = \prod_{n=1}^{N} \left( \frac{\exp(\eta_i + \alpha_n)t_{mn}}{1+\exp(\eta_i + \alpha_n)} \right)^{\delta_{mn}} \quad m\in\{1,\ldots,M\} \label{eq1a1}\tag{1.2}\\[0.15cm]
		& \eta_i \sim \mathcall N(\eta;\mu_i,\sigma^2_\eta) \tag{1.3}
	\end{align}
	where $t_{mn}$ is an entry of the mapping matrix $\mathbf T_{M\times N}$, which indicate how each response category (in rows) is associated to each node (in columns) of the tree. For the right-most tree in Figure \ref{fig2}, the mapping matrix is as follows:
	$$
	\mathbf T_{5 \times 4} = 
	\begin{bmatrix}
		1&&0&&0&&\text{\tiny NA}\\
		1&&0&&1&&\text{\tiny NA}\\
		0&&\text{\tiny NA}&&\text{\tiny NA}&&\text{\tiny NA}\\
		1&&1&&\text{\tiny NA}&&0\\
		1&&1&&\text{\tiny NA}&&1
	\end{bmatrix}
	$$
	with $N=M-1$ being the number of nodes. As $t_{mn}\in\{0,1\}$, $t_{mn}=1$ indicates that the $m$-th category of response involves the node $n$, $t_{mn} = 0$ indicates that the $m$-th category of response does not involve the node $n$, whereas $t_{mn} = \text{\tiny NA}$ indicates that the $m$-th category of response is not connected to the $n$-th node at all. The term $\delta_{mn}$ is defined as follows: $\delta_{mn} = 0$ if $t_{mn}=\text{\tiny NA}$ and $\delta_{mn} = 1$ otherwise. The rater's ability $\eta_i$ is a random quantity from a Normal distribution with mean $\mu_i \in \mathbb R$ and variance $\sigma^2_\eta \in \mathbb R^+$. Usually, $\mu_i = 0$ for most applications, although it can be rewritten as a linear combinations of $K$ variables $\mu_i = \mathbf{x}_i\boldsymbol{\beta}$ to account for the effect of external covariates. Finally, the parameter $\alpha_n \in \mathbb R$ expresses the easiness of choosing the $n$-node of the tree. In general, we may have as many $\alpha$'s as the number of nodes or, more simply, a single $\alpha$ for all the nodes \cite{Boeck_2012}. 
	
	\subsection{About the probabilistic term \eqref{eq1b}} 
	
	The second term of the model is a mixture distribution representing the last two stages of the fuzzy rating process. Conditioned on the first stage $C_i=c$, the final response might be affected by decision uncertainty at some degrees - a case in which $\xi_i > 0$ - or, conversely, it might be free of fuzziness. To exemplify the idea behind this representation, consider once again the right-most decision tree in Figure \ref{fig2}. We expect that a higher degree of decision uncertainty entails a higher difficulty level to navigate the tree structure, which in turn increases all the response probabilities $\boldsymbol{\pi}^y_i$. Conversely, a lower degree of decision uncertainty implies a lower difficulty to go through the tree nodes, which in turn decreases the probability to activate contiguous responses. This suggests to use $\boldsymbol{\pi}^y_i$ in the definition of  \eqref{eq1b}. In particular, we define the mixture probability $\xi_i$ in terms of the normalized Shannon entropy:
	\begin{align}\label{eq1b1}
		\xi_i = -\left({\sum_{m=1}^{M} \pi_{im}^y \ln \pi_{im}^y}\right)\Big/{\ln M} \tag{2.1}
	\end{align}
	and set the mixture components to be Binomial as follows:
	\begin{align}
		& \Prob{L_i = l|C_i,\eta_i; \boldsymbol{\theta}} = \mathcall Bin(l; ~C_i-1, \pi_i^s) \tag{2.2} \label{eq1b2}\\
		& \Prob{R_i = r|C_i,\eta_i; \boldsymbol{\theta}} = \mathcall Bin(r;~ M-C_i, 1-\pi_i^s) \tag{2.3} \label{eq1b3}\\
		& \Prob{L_i = 0|C_i,\eta_i; \boldsymbol{\theta}} = \mathcall Bin(l; ~C_i-1, 0) \tag{2.4} \label{eq1b4}\\
		& \Prob{R_i = 0|C_i,\eta_i; \boldsymbol{\theta}} = \mathcall Bin(r;~ M-C_i, 0) \tag{2.5} \label{eq1b5}
	\end{align}
	where \eqref{eq1b4}-\eqref{eq1b5} are degenerate distribution with mass one on the element zero of the support \cite{cao2012semiparametric}. The parameter $\pi_i^s$ is the probability to activate lower response categories and it is defined as follows:
	\begin{align}
		\pi_i^s =  {\sum_{m\in \{1,\ldots,M\}\setminus C_i} \pi_{im}^y} \Big/ {(1-\pi_{i,m=c_i}^y)} \tag{2.6}\label{eq1b6}
	\end{align}
	under the convention that $\pi_i^s = 0$ if $m=c_i$ and where $\pi_{i,m=c_i}^y$ is the probability of the current response $C_i=c$. Note that the normalized Shannon entropy increases as $\boldsymbol{\pi}^y$ gets uniform and decreases as $\boldsymbol{\pi}^y$ becomes degenerate for a single element of $\{1,\ldots,M\}$. This property makes the entropy measure suitable to quantify varying levels of decision uncertainty in the rating process. 
	
	\subsection{Sampling schema}
	
	In short, the proposed model can be rewritten in terms of the underlying sampling process as follows:
	
	\begin{align}\label{eq2}
		& \eta_i \sim \mathcall N(\eta;\mathbf x_i\boldsymbol{\beta},\sigma^2_\eta)\nonumber \\[0.15cm]
		& C_i^\dagger|\eta_i \sim \mathcall Multinom\big( c^\dagger; 1, \boldsymbol{\pi}^y_i(\boldsymbol{\alpha},\eta_i) \big) \nonumber \\[0.15cm]
		& Z_i|\eta_i \sim \mathcall Bin\big(z;1,\xi_i(\boldsymbol{\alpha},\eta_i)\big)\nonumber\\[0.15cm]
		& Z_i=1~\begin{cases}
			& L_i|C_i,\eta_i \sim \mathcall Bin\big(l;C_i-1,\pi_i^s(\boldsymbol{\alpha},\eta_i)\big) \\
			& R_i|C_i,\eta_i \sim \mathcall Bin\big(l;M-C_i,1-\pi_i^s(\boldsymbol{\alpha},\eta_i)\big)
		\end{cases}\\[0.20cm]
		& Z_i=0~\begin{cases}
			& L_i|C_i,\eta_i \sim \mathcall Bin\big(l;C_i-1,0\big)\nonumber\\
			& R_i|C_i,\eta_i \sim \mathcall Bin\big(l;M-C_i,0\big)\nonumber
		\end{cases}
	\end{align}
	where $C_i = \mathcall I(C_i^\dagger)$, $\boldsymbol{\pi}^y_i(\boldsymbol{\alpha},\eta_i)$ is defined via Eq. \eqref{eq1a1}, $\xi_i(\boldsymbol{\alpha},\eta_i)$ is defined via Eq. \eqref{eq1b1}, whereas $\pi_i^s(\boldsymbol{\alpha},\eta_i)$ is defined according to Eq. \eqref{eq1b6}. 
	
	According to the stage-wise representation of the rating response process, model \eqref{eq2} is self-consistent in the manner through which the fuzzy data $\mathbf{\tilde y}$ are modeled. Indeed, given an IRTree structure according to which the rating process is supposed to behave, a particular instance of $\boldsymbol{\theta} = \{\boldsymbol{\alpha}, \boldsymbol{\beta}, \sigma^2_\eta\}$ gives rise to a cascade computations from the input to the output $\{\hat c, \hat l, \hat r\}$ through the model equations. As a result, external information in terms of explaining variables or covariates can be plugged-in to the model through the model parameters only and there is no way to link them to the outcome variable directly. \\
	Finally, the probability of a fuzzy response is as follows:
	\begin{align}\label{eq3}
		\Prob{Y_i =& (c_i,l_i,r_i)|\eta_i;\boldsymbol{\theta}} = ~ \pi_{i,m=c_i}^y \times \nonumber\\ 
		& \times \Big[ \xi_i \binom{c_i-1}{l_i}\binom{M-c_i}{r_i} (\pi_i^s)^{l_i+M-c_i-r_i} \cdot (1-\pi_i^s)^{r_i+c_i-l_i-1} ~+ \nonumber\\[0.15cm]
		& + (1-\xi_i) \binom{c_i-1}{l_i}\binom{M-c_i}{r_i} 0^{l_i+r_i} \cdot 1^{M-r_i-l_i-1} \Big] \times \\
		& \times \frac{1}{\sigma_\eta\sqrt{2\pi}}\exp\left(-\frac{1}{2\sigma^2_\eta}(\eta_i-\mathbf x_i\boldsymbol{\beta})^2\right) \nonumber
	\end{align}
	where $\pi_{i,m=c_i}^y$ indicates the probability of the response $C_i=c$.
	
	\subsection{Parameter estimation}
	
	Model \eqref{eq3} implies the following parameters $\boldsymbol{\theta} = \{\boldsymbol{\alpha}, \boldsymbol{\beta}, \sigma^2_\eta\} \subset \mathbb R^N\times \mathbb R^K \times \mathbb R_+$. Since the model uses a logistic function to determine $\boldsymbol{\pi}^y$, we can further simplify the parameter estimation by restricting the parameter space in a subset of reals, for instance by means of the following constraints: $|(\boldsymbol{\alpha},\boldsymbol{\beta})|^T\mathbf 1_{N+K} \leq 5$ and $\sigma_\eta \in (0,3.5]$. They are justified by the simple fact that the logistic curve increases quickly only in a small subset of its domain. The model parameters can be estimated by maximizing the marginal likelihood function, which is obtained by integrating out the random terms $\eta_1,\ldots,\eta_I$ from the full likelihood function \cite{pawitan2001all}. This requires the computation of the following marginal probability distribution:
	\begin{align}\label{eq4}
		\Prob{Y_i = (c_i,l_i,r_i);\boldsymbol{\theta}} & = \bigintssss_{\mathbb R} \Prob{Y_i = (c_i,l_i,r_i)|\eta_i;\boldsymbol{\alpha}}f_{\eta_i}(\eta;\mathbf{x}_i\boldsymbol{\beta},\sigma_\eta^2)~d\eta \nonumber\\
		& \propto \bigintssss_{\mathbb R} \pi_{i,m=c_i}^y \Big[ \xi_i\Big( (\pi_i^s)^{l_i+M-c_i-r_i} \cdot (1-\pi_i^s)^{r_i+c_i-l_i-1} - \nonumber \\
		& \hspace{0.5cm} - 0^{l_i+r_i} \cdot 1^{M-r_i-l_i-1} \Big) + 0^{l_i+r_i} \cdot 1^{M-r_i-l_i-1} \Big] \times \\
		& \hspace{0.5cm} \times \exp\left(-\frac{1}{2\sigma^2_\eta}(\eta_i-\mathbf x_i\boldsymbol{\beta})^2\right) ~d\eta \nonumber \\
		& \propto \bigintssss_{\mathbb R} h(c_i,l_i,r_i,\boldsymbol{\alpha},\eta_i) \exp\left(-\frac{1}{2\sigma^2_\eta}(\eta_i-\mathbf x_i\boldsymbol{\beta})^2\right) ~d\eta \nonumber
	\end{align}
	where the integral can be solved numerically via the Gauss-Hermite quadrature. By the change of variable $d_i = \sigma_\eta\sqrt{2}\eta_i + \mathbf{x}_i\boldsymbol{\beta}$, the integral is approximated as follows:
	\begin{align}\label{eq5}
		\Prob{Y_i = (c_i,l_i,r_i);\boldsymbol{\theta}} & \propto \bigintssss_{\mathbb R} h(c_i,l_i,r_i,\boldsymbol{\alpha},d_i) \exp\big(-\frac{1}{2\sigma^2_\eta}d_i^2\big) ~dd_i \nonumber\\
		& \approx \frac{1}{\sqrt{\pi}} ~ \sum_{h=1}^{H} h(c_i,l_i,r_i,\boldsymbol{\alpha},\sigma_\eta\sqrt{2}\gamma_h+\mathbf x_i\boldsymbol{\beta}) ~\omega_h
	\end{align}
	where $\gamma_1,\ldots,\gamma_H$ and $\omega_1,\ldots,\omega_H$ are the nodes and weights of the quadrature to be computed numerically for a fixed $H$ \cite{golub1969calculation}. Finally, the log-likelihood function:
	\begin{align}\label{eq6}
		\ln \mathcall L(\boldsymbol{\theta}) \propto \sum_{i=1}^{I} \ln \left( \sum_{h=1}^{H} h(c_i,l_i,r_i,\boldsymbol{\alpha},\sigma_\eta\sqrt{2}\gamma_h+\mathbf x_i\boldsymbol{\beta}) ~\omega_h \right)
	\end{align}
	can be maximized numerically via either the Broyden-Fletcher-Goldfarb-Shanno (BFGS) or the Augmented Lagrangian (AUGLAG) algorithms. Note that in the first case the variance parameter has to be transformed to lie into the real line (e.g., via $\exp$ function) whereas in the second case the constraints $|(\boldsymbol{\alpha},\boldsymbol{\beta})|^T\mathbf 1 \leq 5$ and $\sigma_\eta \in (0,3.5]$ can be directly plugged in to the optimization routine.

\section{Application}

In this section we illustrate the characteristics of the proposed model by means of an application to a real dataset. In particular, data refers to a survey administered to $n=69$ young drivers in Trentino region (north-est of Italy). Of these, 45\% were women with mean age of 18.23 years. All participants were young drivers with an average of driving experience of 12 months since receipt of their driver's license. About 74\% of them drove frequently during the week, 26\% drove once a week. Participants were asked to self-assess their reckless-driving behavior (\texttt{RDB}) along with a short version of the Driving Anger Scale (\texttt{DAS}), adopted to evaluate the driving anger provoked by someone else's behaviors. Ratings were collected using a four-point direct fuzzy rating scale (see Figure \ref{fig1}). For both scales, higher categories indicate higher scores on \texttt{RDB} and \texttt{DAS} items, respectively. To simplify the interpretation of the results, the items of the Driving Anger Scale were aggregated to form a crisp total score. In the next data analysis, the fuzzy variable \texttt{RDB} was used as response variable whereas the \texttt{DAS} total score was used as crisp predictor. 

\begin{table}[h!]
	\centering
	\begin{tabular}{c|c|c|c|c}
		\hline\hline
		Model & Covariates & No. of parameters & $\ln\mathcall L(\boldsymbol{\theta})$ & BIC \\ 
		\hline
		M1: linear tree & - & 2 & -161.15 & 330.767 \\ 
		M2: linear tree & \texttt{sex} & 3 & -157.855 & 328.412 \\ 
		\rowcolor{gray} M3: linear tree & \texttt{sex}, \texttt{DAS} & 4 & -155.268 & 327.472 \\ 
		M4: linear tree & \texttt{sex}, \texttt{DAS}, \texttt{sex:DAS} & 5 & -155.253 & 331.676 \\ 
		M5: nested tree & \texttt{sex}, \texttt{DAS} & 5 & -158.937 & 339.044 \\ 
		\hline\hline
	\end{tabular}
	\caption{Application: Models for the RDB fuzzy rating data. Note that model M3 is the best model according to the lowest BIC criterion.} 
	\label{tab1}
\end{table}

\begin{table}[h!]
	\centering
	\begin{tabular}{c|c|c}
		\hline\hline
		Parameter & Estimate & Std. Error \\ 
		\hline
		$\alpha$ & -1.248 & 0.09 \\ 
		$\beta_{\texttt{sex}}$ & 0.408 & 0.119 \\ 
		$\beta_{\texttt{das}}$ & 1.284 & 0.093 \\ 
		$\sigma_\eta$ & 0.005 & 19.947 \\ 
		\hline\hline
	\end{tabular}
	\caption{Application: Parameter estimation and standard errors for the model M3.} 
	\label{tab2}
\end{table}

Three models (M1-M4) with a linear decision tree (see Figure \ref{fig2}, leftmost panel) and an additional model (M5) with a nested decision tree structure (see Figure \ref{fig2}, rightmost panel) were run on \texttt{RDB}. The models varied in terms of covariates (see Table \ref{tab1}). In particular, model M1 involved no covariates and a common $\alpha$ parameter for all the $N=4$ nodes of the decision tree. On the contrary, models M2-M4 differed from M1 just in terms of covariates, with M4 including the interaction term \texttt{sex:DAS}. Finally, model M5 differed from M3 as this uses a different decision tree with a nested structure (see Figure \ref{fig2}, rightmost panel). The final model was chosen according to the fitting measure $BIC = -2\ln\mathcall L(\boldsymbol{\theta}) + p\ln I$, with $p$ being the number of parameters implied by the model. The best model is that achieving the lowest BIC, in this case M3. Table \ref{tab2} reports the estimated parameters whereas Figures \ref{fig4}-\ref{fig5} show the marginal effects for the chosen model. As it includes the categorical covariate \texttt{sex}, the parameters ${\alpha}$ codify the intercept of the model across all the nodes, which in this case is the coefficient for the level \texttt{sex=F} when \texttt{DAS=0}. 

\begin{figure}[h!]
	\hspace{-0.75cm}
	\resizebox{16cm}{!}{\input{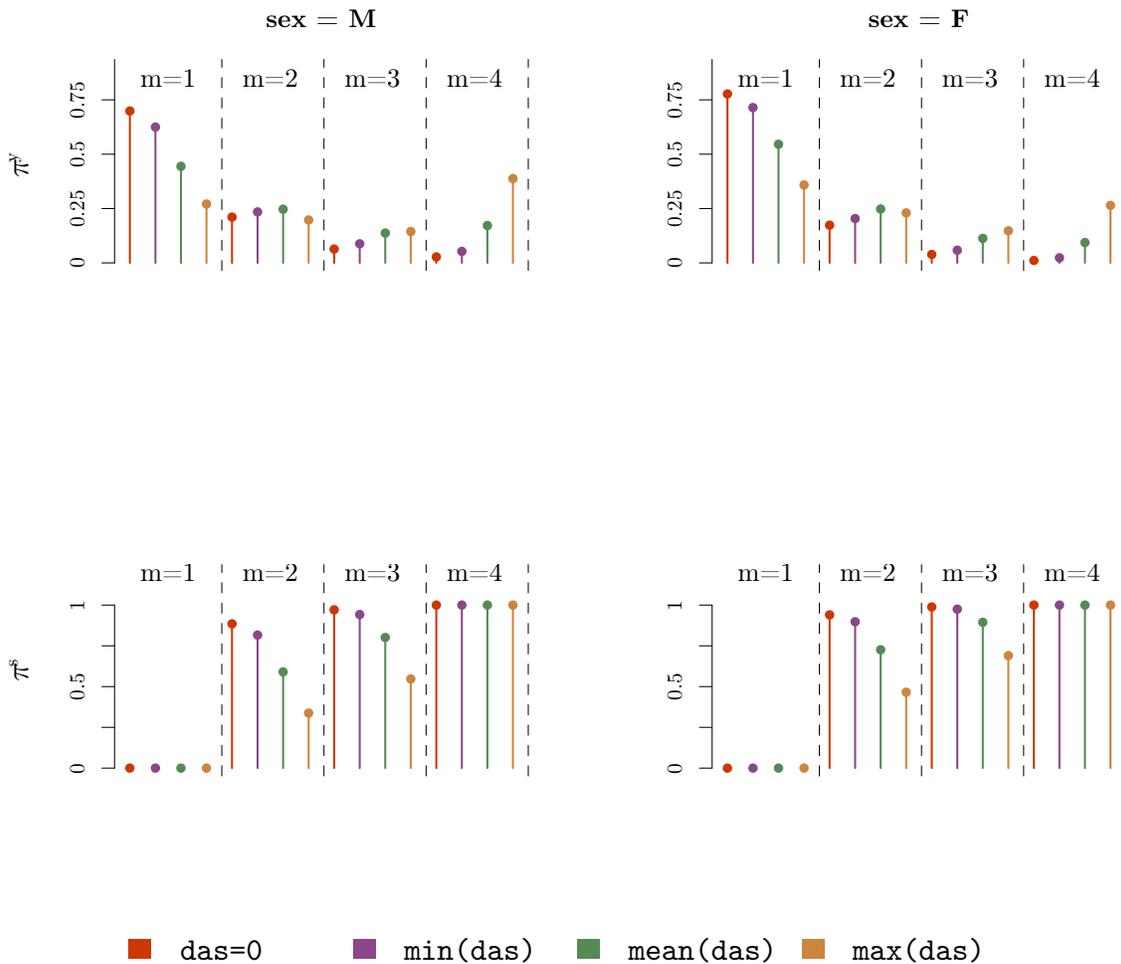}}
	\caption{Application: Marginal effects computed over four reference values of \texttt{das} (\texttt{das=0}, \texttt{das=min}, \texttt{das=mean}, \texttt{das=max}) and for both \texttt{sex=M} and \texttt{sex=F}. The effects are computed for the response probability $\boldsymbol{\pi}^y$ (first row) and for the probability to activate a lower response $\pi^s$ (second row). Note that $m=1,\ldots,m=4$ indicate the response categories of the rating scale. }
	\label{fig4}	
\end{figure}

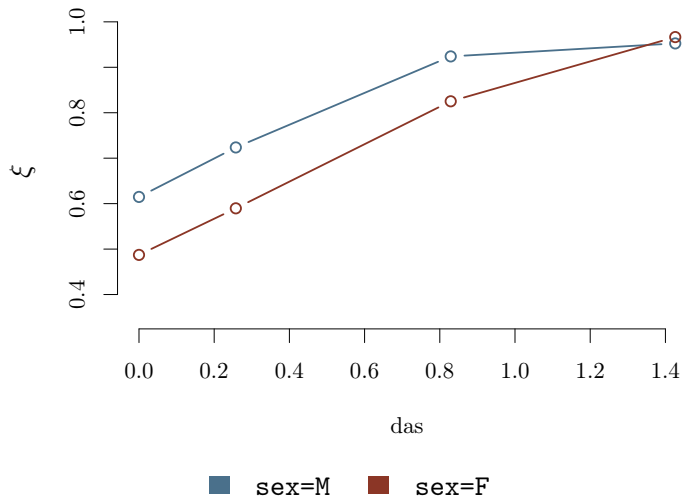
\begin{figure}[h!]
	\centering
	\resizebox{10cm}{!}{
\begin{tikzpicture}[x=1pt,y=1pt]
\definecolor{fillColor}{RGB}{255,255,255}
\path[use as bounding box,fill=fillColor,fill opacity=0.00] (0,0) rectangle (325.21,252.94);
\begin{scope}
\path[clip] ( 61.43, 83.11) rectangle (314.37,220.42);
\definecolor{drawColor}{RGB}{74,112,139}

\path[draw=drawColor,line width= 0.8pt,line join=round,line cap=round] ( 76.14,143.52) -- (107.75,159.69);

\path[draw=drawColor,line width= 0.8pt,line join=round,line cap=round] (118.62,164.77) -- (201.42,199.87);

\path[draw=drawColor,line width= 0.8pt,line join=round,line cap=round] (212.94,202.56) -- (299.02,207.55);

\path[draw=drawColor,line width= 0.8pt,line join=round,line cap=round] ( 70.80,140.79) circle (  2.25);

\path[draw=drawColor,line width= 0.8pt,line join=round,line cap=round] (113.09,162.42) circle (  2.25);

\path[draw=drawColor,line width= 0.8pt,line join=round,line cap=round] (206.95,202.21) circle (  2.25);

\path[draw=drawColor,line width= 0.8pt,line join=round,line cap=round] (305.01,207.89) circle (  2.25);
\end{scope}
\begin{scope}
\path[clip] (  0.00,  0.00) rectangle (325.21,252.94);
\definecolor{drawColor}{RGB}{0,0,0}

\path[draw=drawColor,line width= 0.4pt,line join=round,line cap=round] ( 70.80, 83.11) -- (300.72, 83.11);

\path[draw=drawColor,line width= 0.4pt,line join=round,line cap=round] ( 70.80, 83.11) -- ( 70.80, 77.11);

\path[draw=drawColor,line width= 0.4pt,line join=round,line cap=round] (103.64, 83.11) -- (103.64, 77.11);

\path[draw=drawColor,line width= 0.4pt,line join=round,line cap=round] (136.49, 83.11) -- (136.49, 77.11);

\path[draw=drawColor,line width= 0.4pt,line join=round,line cap=round] (169.33, 83.11) -- (169.33, 77.11);

\path[draw=drawColor,line width= 0.4pt,line join=round,line cap=round] (202.18, 83.11) -- (202.18, 77.11);

\path[draw=drawColor,line width= 0.4pt,line join=round,line cap=round] (235.02, 83.11) -- (235.02, 77.11);

\path[draw=drawColor,line width= 0.4pt,line join=round,line cap=round] (267.87, 83.11) -- (267.87, 77.11);

\path[draw=drawColor,line width= 0.4pt,line join=round,line cap=round] (300.72, 83.11) -- (300.72, 77.11);

\node[text=drawColor,anchor=base,inner sep=0pt, outer sep=0pt, scale=  1.00] at ( 70.80, 61.51) {0.0};

\node[text=drawColor,anchor=base,inner sep=0pt, outer sep=0pt, scale=  1.00] at (103.64, 61.51) {0.2};

\node[text=drawColor,anchor=base,inner sep=0pt, outer sep=0pt, scale=  1.00] at (136.49, 61.51) {0.4};

\node[text=drawColor,anchor=base,inner sep=0pt, outer sep=0pt, scale=  1.00] at (169.33, 61.51) {0.6};

\node[text=drawColor,anchor=base,inner sep=0pt, outer sep=0pt, scale=  1.00] at (202.18, 61.51) {0.8};

\node[text=drawColor,anchor=base,inner sep=0pt, outer sep=0pt, scale=  1.00] at (235.02, 61.51) {1.0};

\node[text=drawColor,anchor=base,inner sep=0pt, outer sep=0pt, scale=  1.00] at (267.87, 61.51) {1.2};

\node[text=drawColor,anchor=base,inner sep=0pt, outer sep=0pt, scale=  1.00] at (300.72, 61.51) {1.4};

\path[draw=drawColor,line width= 0.4pt,line join=round,line cap=round] ( 61.43, 98.13) -- ( 61.43,217.32);

\path[draw=drawColor,line width= 0.4pt,line join=round,line cap=round] ( 61.43, 98.13) -- ( 55.43, 98.13);

\path[draw=drawColor,line width= 0.4pt,line join=round,line cap=round] ( 61.43,117.99) -- ( 55.43,117.99);

\path[draw=drawColor,line width= 0.4pt,line join=round,line cap=round] ( 61.43,137.86) -- ( 55.43,137.86);

\path[draw=drawColor,line width= 0.4pt,line join=round,line cap=round] ( 61.43,157.73) -- ( 55.43,157.73);

\path[draw=drawColor,line width= 0.4pt,line join=round,line cap=round] ( 61.43,177.59) -- ( 55.43,177.59);

\path[draw=drawColor,line width= 0.4pt,line join=round,line cap=round] ( 61.43,197.46) -- ( 55.43,197.46);

\path[draw=drawColor,line width= 0.4pt,line join=round,line cap=round] ( 61.43,217.32) -- ( 55.43,217.32);

\node[text=drawColor,rotate= 90.00,anchor=base,inner sep=0pt, outer sep=0pt, scale=  1.00] at ( 47.03, 98.13) {0.4};

\node[text=drawColor,rotate= 90.00,anchor=base,inner sep=0pt, outer sep=0pt, scale=  1.00] at ( 47.03,137.86) {0.6};

\node[text=drawColor,rotate= 90.00,anchor=base,inner sep=0pt, outer sep=0pt, scale=  1.00] at ( 47.03,177.59) {0.8};

\node[text=drawColor,rotate= 90.00,anchor=base,inner sep=0pt, outer sep=0pt, scale=  1.00] at ( 47.03,217.32) {1.0};
\end{scope}
\begin{scope}
\path[clip] (  0.00,  0.00) rectangle (325.21,252.94);
\definecolor{drawColor}{RGB}{0,0,0}

\node[text=drawColor,anchor=base,inner sep=0pt, outer sep=0pt, scale=  1.00] at (187.90, 37.51) {das};

\node[text=drawColor,rotate= 90.00,anchor=base west,inner sep=0pt, outer sep=0pt, scale=  1.25] at ( 23.03,149.13) {$\xi$};
\end{scope}
\begin{scope}
\path[clip] ( 61.43, 83.11) rectangle (314.37,220.42);
\definecolor{drawColor}{RGB}{139,54,38}

\path[draw=drawColor,line width= 0.8pt,line join=round,line cap=round] ( 76.20,118.06) -- (107.69,133.21);

\path[draw=drawColor,line width= 0.8pt,line join=round,line cap=round] (118.46,138.49) -- (201.58,179.92);

\path[draw=drawColor,line width= 0.8pt,line join=round,line cap=round] (212.71,184.24) -- (299.24,209.01);

\path[draw=drawColor,line width= 0.8pt,line join=round,line cap=round] ( 70.80,115.45) circle (  2.25);

\path[draw=drawColor,line width= 0.8pt,line join=round,line cap=round] (113.09,135.81) circle (  2.25);

\path[draw=drawColor,line width= 0.8pt,line join=round,line cap=round] (206.95,182.59) circle (  2.25);

\path[draw=drawColor,line width= 0.8pt,line join=round,line cap=round] (305.01,210.66) circle (  2.25);
\end{scope}
\begin{scope}
\path[clip] (  0.00,  0.00) rectangle (325.21,252.94);
\definecolor{fillColor}{RGB}{74,112,139}

\path[fill=fillColor] (101.29, 18.75) rectangle (110.29, 11.25);
\definecolor{fillColor}{RGB}{139,54,38}

\path[fill=fillColor] (171.04, 18.75) rectangle (180.04, 11.25);
\definecolor{drawColor}{RGB}{0,0,0}

\node[text=drawColor,anchor=base west,inner sep=0pt, outer sep=0pt, scale=  1.25] at (121.54, 10.70) {\texttt{sex=M}};

\node[text=drawColor,anchor=base west,inner sep=0pt, outer sep=0pt, scale=  1.25] at (191.29, 10.70) {\texttt{sex=F}};
\end{scope}
\end{tikzpicture}}
	\caption{Application: Probability $\xi$ to activate the mixture components (marginal effect) computed as a function of \texttt{das} and for both \texttt{sex=M} and \texttt{sex=F}.}
	\label{fig5}	
\end{figure}

Overall, when \texttt{DAS=0}, participants in the group \texttt{sex=F} showed a stronger tendency to choose lower response categories ($\hat\alpha=-1.248$, $\sigma_{\hat\alpha}=0.09$) if compared to participants in the group \texttt{sex=M} ($\hat\beta_{\texttt{sex}}=0.408$, $\sigma_{\hat\beta_{\texttt{sex}}}=0.119$). Similarly, \texttt{DAS} was positively associated to \texttt{RDB} ($\hat\beta_{\texttt{das}}=1.284$, $\sigma_{\hat\beta_{\texttt{das}}}=0.093$) and acted by increasing the tendency to activate the last nodes of the decision tree (see Figure \ref{fig4}, first row). With regards to the parameter $\xi$, participants in the group \texttt{sex=M} showed a higher probability to activate the spread components of the fuzzy response across all the levels of \texttt{DAS} as opposed to participants in the group \texttt{sex=F} (see Figure \ref{fig5}). Thus, all in all, the results suggest that driving anger increased the levels of decision uncertainty, with male participants showing a larger fuzziness if compared to female participants.

\section{Conclusions}

In this contribution we have described a new statistical model for fuzzy rating responses that are collected by means of direct fuzzy rating scales. With the aim of representing the stage-wise decision process underlying a rating response, the model revolves around the adoption of a conditional representation where a Multinomial tree component is coupled with a mixture of Binomial distributions to represent the fuzziness of rating responses. A nice advantage of the proposed method is its ability to deal with LR-type triangular fuzzy data in terms of the stage-wise mechanisms supposed to drive the unobserved rating response process. However, as for any statistical model, it has some limitations. In particular, the current version of the model does not take into account the shape of LR-type fuzzy numbers and it cannot be used in a multivariate context (i.e., the model works with a single outcome variable per time). Further investigations might consider these limitations more explicitly, for instance by means of additional simulation studies. The results of this contribution should be considered as initial findings to the problem of analysing fuzzy responses in a psychometric modeling context.

\bibliographystyle{plain}
\bibliography{biblio}

\end{document}